\documentclass[]{aa}
\usepackage{graphicx}

\begin{document}

\title{A broadband leptonic model for gamma-ray emitting microquasars}
\authorrunning{Bosch-Ramon, Romero \& Paredes}
\titlerunning{Broad-band emission from microquasars}

\author{V. Bosch-Ramon\inst{1} \and G. E. Romero\inst{2,3,}\footnote{Member of CONICET} 
\and J.~M. Paredes\inst{1}}

\institute{Departament d'Astronomia i Meteorologia, Universitat de Barcelona, Av. 
Diagonal 647, 08028 Barcelona, Catalonia, Spain; vbosch@am.ub.es, jmparedes@ub.edu.
\and Instituto Argentino de Radioastronom\'{\i}a, C.C.5,
(1894) Villa Elisa, Buenos Aires, Argentina; romero@iar.unlp.edu.ar.
\and Facultad de Ciencias Astron\'omicas y Geof\'{\i}sicas, UNLP, 
Paseo del Bosque, 1900 La Plata, Argentina.}

\offprints{V. Bosch-Ramon \\ \email{vbosch@am.ub.es}}

\abstract{
Observational and theoretical studies point to microquasars (MQs) as possible counterparts of a
significant  fraction of the unidentified gamma-ray sources detected so far. At present, 
a proper scenario to explain the emission beyond soft X-rays from these
objects is not known, nor what the precise connection is between the radio and the high-energy radiation. We
develop a new model where the MQ jet is dynamically dominated  by cold protons and radiatively
dominated by relativistic leptons. The matter content and power of the jet are both related with the
accretion process. The magnetic field is assumed to be close to equipartition, although it is
attached to and dominated by the jet matter. For the relativistic particles in the jet, their
maximum  energy depends on both the acceleration efficiency and the energy losses. The model takes into
account the interaction of the relativistic jet particles 
with the magnetic field and all the photon and matter fields. Such interaction 
produces significant amounts of radiation from radio to very high
energies through synchrotron, relativistic Bremsstrahlung, and inverse Compton (IC) processes.
Variability of the emission produced by changes in the  accretion process (e.g. via orbital
eccentricity) is also expected. The effects of the gamma-ray absorption by the external photon fields
on  the gamma-ray spectrum have been taken into account, revealing clear spectral  features that might
be observed. This model is consistent to the accretion
scenario, energy conservation laws, and current observational knowledge, and can provide deeper
physical information of the source when tested against multiwavelength data.
}

\maketitle

\section{Introduction} \label{intro}

Microquasars are X-ray binary systems (XRBs) with
relativistic bipolar outflows or jets (Mirabel \& Rodr{\'{\i}}guez \cite{Mirabel99}). These extended
structures have been observed in galactic objects at radio wavelengths from the seventies (SS~433;
Spencer \cite{Spencer79}, Hjellming \& Johnston \cite{Hjellming81}). The inner region of the disk emits
thermally at soft X-rays, losing accretion kinetic energy through viscosity-related phenomena.
Additionally, there seems to be evidence supporting the existence of a hot relativistic plasma around
the compact object (called `corona' hereafter), which is radiatively significant
during the so-called low-hard state of the source. Corona electrons would Comptonize soft disk photons
to higher energies (McClintock \& Remillard \cite{McClintock04}). Transport of angular momentum and
kinetic energy linked to a specific inner disk magnetic field configuration could lead to the formation
of a jet (Meier \cite{Meier03}).

The detection of extended non-thermal radio emission provided clear evidence for the presence of relativistic leptons
in the jets of MQs, although it was not considered in general that jets could emit significantly at X-rays or beyond. 
Paredes et~al. (\cite{Paredes00}) proposed the microquasar LS~5039 as the counterpart of the EGRET source
3EG~J1824$-$1514 (Hartman et~al. \cite{Hartman99}). In their scenario, the jet relativistic electrons scatter the
photons of the massive stellar companion, showing that microquasar jets are possible sources of gamma-rays.  Further
statistical and theoretical studies showed that microquasars could be behind some of the unidentified gamma-ray sources
in the Galaxy (Kaufman Bernad\'o et~al. \cite{Kaufman02}, Romero et~al. \cite{Romero04}, Bosch-Ramon et~al.
\cite{Bosch05a}, \cite{Bosch05b}). Observational evidence of jets as high energy emitters came from the detection of
X-ray extended emission (e.g., in SS~433, Marshall et~al. \cite{Marshall02},  Migliari et~al. \cite{Migliari02};
XTE~J1550-564, Corbel et~al. \cite{Corbel02}; and H 1743-322, Corbel et al. \cite{Corbel05}). The recent detection by
the ground-based Cherenkov telescope HESS of TeV emission coming from the microquasar LS~5039 (Aharonian et~al. 2005)
largely confirms the association proposed by Paredes et~al. (\cite{Paredes00}), and it is strong evidence that
microquasars are sources of very high-energy gamma-rays, their jets being the best candidates to generate such
emission.

For the modeling of gamma-ray emission from jets of MQs, there have been two types of approach. One considers that
hadrons lead radiative processes at GeV-TeV gamma-rays and beyond  (Romero et~al. \cite{Romero03}, \cite{Romero05},
Romero \& Orellana \cite{Romero-Orellana05}),  producing detectable amounts of neutrinos  (Torres et~al.
\cite{Torres04}), and leaving electrons as possible significant emitters only at lower energies. These are the
so-called hadronic models. The other approach extends the energy of leptons from synchrotron radio emitting energies
exploring  inverse Compton and/or synchrotron emission in the jets (i.e. Atoyan \& Aharonian  \cite{Atoyan99}, Markoff
et~al. \cite{Markoff01},  Georganopoulos et~al. \cite{Georganopoulos02}, Bosch-Ramon et~al. \cite{Bosch05a}).  These
are the so-called leptonic models.  All these models developed so far are important to investigate to what extent each
mechanism of emission would be relevant under different circumstances, and what  level of physical detail is required
for realistic modeling with the available observational data. Nevertheless, a comprehensive MQ jet model attempting to
explain emission properties in the whole range of spectral frequencies, in accordance with the energy and matter
constraints imposed by accretion and the conservation energy law at the microscopic level, is still lacking. 

In this paper, we investigate persistent MQ compact jets to give multiwavelength
and variability predictions consistent with the MQ scenario as a whole. We use the term `compact jet'
or `jet' referring to the type of outflows thought to be present during the low-hard state
(Fender et~al. \cite{Fender03a}). The extended radio lobes, which are also observed in some MQs (e.g.
1E~1740.7$-$2942; Mirabel et~al. \cite{Mirabel92}), and the blobs ejected during state transitions  (e.g.
GRS~1915+105, Mirabel et~al. \cite{Mirabel98}) are not considered as compact jets.  
`Consistent' means here to develop the model taking into account the total amount of matter available for
accretion, the pressure relationship between compact jets and their environments, the standard models for
accretion and jet ejection, the mechanism  for particle acceleration, the pair creation and annihilation
rates and the law of microscopic energy conservation. Semi-analytical calculations have been
implemented to compute all the significant emission and absorption mechanisms that take place in the jet:
synchrotron, relativistic Bremsstrahlung with internal and external matter fields, inverse Compton with
internal and external photon fields, and creation and annihilation of pairs. In a forthcoming paper, we 
will present an application of the present model to the microquasar LS~5039 (Paredes et~al. \cite{Paredes05}).

In Section~\ref{pict}, the general picture of the model is presented; in Section~\ref{jet}, the details of the
jet model are given; in Section~\ref{rad}, the considered radiative processes are explained. The resulting
spectral energy distributions (SEDs) for several relevant situations are shown and discussed in
Section~\ref{seds}, as are the variability properties  of the model. Other questions implied by our model
are discussed in Section~\ref{disc}, and all the treated issues are summarized in Section~\ref{sum}.

\section{General picture}\label{pict}

The MQ scenario considered here consists of a binary system formed by a star, either of low or  high mass,
and a compact object, either a black hole or a neutron star. At this stage, the nature of the compact
object is not relevant. The stellar companion feeds the accretion process onto the compact object. Part of
the energy associated with the accreted matter is radiated in an accretion disk, and part is converted to
kinetic and magnetic energy of the accretion flow under the effects of the compact object potential well.
In the low-hard state, when the accretion rate is moderately low, the accretion disk is well described 
by a standard optically thick and geometrically thin
Keplerian disk (Shakura \& Sunyaev \cite{Shakura73}) up to a certain transition
radius, assumed here to be $R_{\rm disk}\sim50R_{\rm Sch}$. 
At distances to the compact object smaller than the transition radius, we assume the existence of a hot plasma 
(the corona), whose relativistic electrons scatter soft disk photons from the geometrically thin disk.
The properties of this inner region are considered to be suitable for jet ejection phenomena to take
place (Meier \cite{Meier03}). The energetics of the jet is assumed to be dominated by accretion,
and further energy sources like compact object rotation have been neglected (an approximation that seems to
be reasonable, see Hujeirat \cite{Hujeirat04}). A more extended discussion on jet formation and energetics is
present in Sect.~\ref{jetf}.

If the jet is formed by accreted matter, it will contain protons and electrons, as well as a
magnetic field ($B$) associated with the plasma. Our assumption is that the  matter kinetic luminosity is higher
than the magnetic luminosity (or total magnetic energy crossing a jet cross-section per time unit) in the
jet regions we are concerned with, although the magnetic field can be still significant once the jet is
formed, since the ejection mechanism is likely to be magneto-hydrodynamical. For simplicity, the
jet is supposed to be perpendicular to the orbital plane. An important fact is that variability can be
easily reproduced when the accretion rate is not constant due to, e.g., orbital eccentricity. 
If protons are relativistic and a confining mechanism is absent, a mildly relativistic jet (see
Sect.~\ref{content}) will expand  relativistically, appearing not collimated. 
Therefore, cold protons instead of relativistic
protons should dominate the jet pressure, with free expansion speeds smaller than in a full
relativistic jet, allowing for a small jet opening angle.
The shock acceleration mechanism,
assumed at this stage to take place all along the jet due to  velocity variations in the ejected matter
(Rees \cite{Rees78}), can effectively accelerate particles, but must such particles diffuse
through the shock. This condition is fulfilled only by a small fraction of protons that, assuming they follow a
thermal distribution, populate the highest energy tail. The condition that particles must diffuse
through the shock to be effectively accelerated imposes a more restrictive condition on the electrons since,
in general, they have a diffusion mean free path much smaller than protons (i.e. lower temperatures), it being
unlikely that shock acceleration will operate significantly on any fraction of the electron thermal
distribution that also forms the jet. Therefore, we assume that an unspecified injection mechanism, like
some kind of plasma instability, operates on the jet electrons heating them enough so they diffuse through the
shock (see Sect.~\ref{content}).

\begin{figure}
\resizebox{\hsize}{!}{\includegraphics{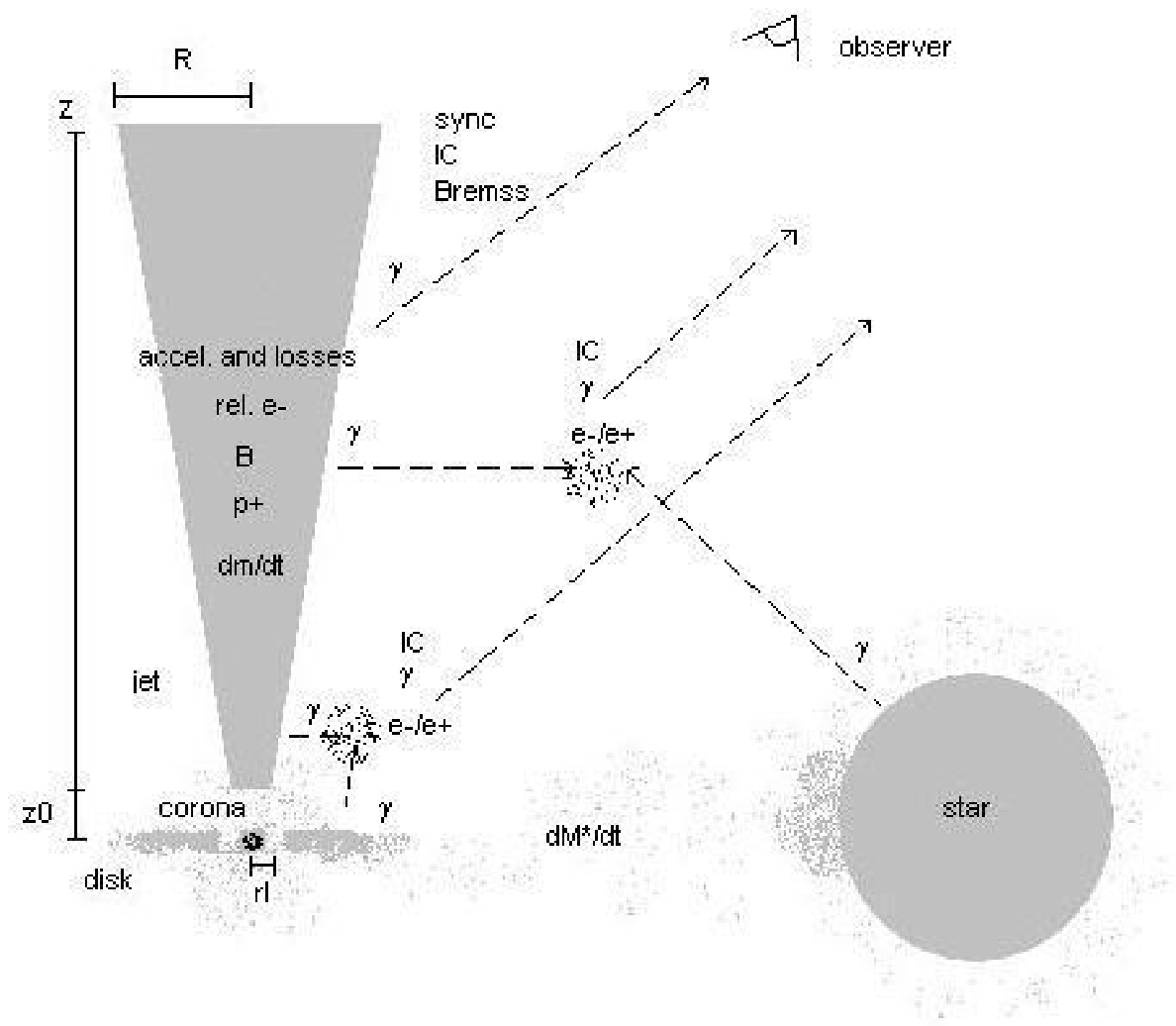}}
\caption{Picture of the microquasar scenario. The main features considered by the model are listed in their 
associated region: accretion and jet matter rate, magnetic field, particle acceleration, energy losses, 
leptonic emission, pair creation and secondary emission. The plot is not to scale.}
\label{fig0}
\end{figure}

In this scenario, radiative processes that take place in the jet can produce significant emission in the
whole spectrum (see Sects.~\ref{rad} and \ref{seds}). Although there could still be enough protons to be
significant from the radiative point of view, treatments on them can be found elsewhere (Romero et~al.
\cite{Romero03}, \cite{Romero05}), and we will focus on the leptonic component only\footnote{We note that
the total luminosity radiated by relativistic protons through proton-proton interaction should be close to
that calculated for relativistic Bremsstrahlung due to similar cross-sections and target densities, assuming
the same energy distribution for both relativistic protons and relativistic electrons}. 
Synchrotron emission can be important from radio to soft gamma-rays, if the
magnetic field is close to equipartition with matter. Moreover, in the regions close to the jet
ejection point, magnetic and relativistic particle densities are high enough for synchrotron self-Compton
process (SSC) to be considered. Moreover, other self-Compton processes can occur, although SSC will be
dominant in most cases. IC interactions between jet electrons and external photon fields (external
Compton or EC) can also be significant. If the companion star is massive, stellar IC scattering can be
dominant at the highest energies. EC scattering with disk and corona photon fields could be non-negligible
as well, and relevant for a low-mass system (see, nonetheless, Grenier et~al. \cite{Grenier05}). All IC
fields appear to dominate emission from soft gamma-rays to very high-energy gamma-rays (VHE), although a
relativistic Bremsstrahlung component can also reach very high energies, generally being a minor
component.  Besides radiative processes, absorption processes are to be taken into account regarding
gamma-ray radiation created in the jet. Absorption of photons by pair creation can be significant at the
base of the jet, and also significant within the binary system if a massive and very luminous star is
present. Moreover, secondaries created by electromagnetic cascades should be considered, since their
contribution is not negligible. In this work, we provide a rough estimate of these effects on the observed
spectrum. It is discussed in Sect.~\ref{rad} and Sect.~\ref{seds}. 
For clarity, in Fig.~\ref{fig0}, we show a rough
sketch of the microquasar scenario.

Four different cases are explored in this paper. First, a high-mass and a low-mass XRB presenting a jet
with a high particle acceleration efficiency are studied in case {\bf A} and {\bf B}, respectively. For
these two particular scenarios, the accretion disk and the corona have been taken to be faint. Here, the
effect of the star on the predicted spectra is investigated. In case {\bf C}, the model is
applied to a high-mass system considering luminous accretion disk and corona with a low particle
acceleration efficiency. For these three cases the velocity is fixed to a mildly relativistic
value. This provides restrictions to the fraction of accreted matter that is ejected, since the energy
budget is limited and some part is radiated during the accretion process. Also, the different acceleration
efficiency has a strong effect on the spectrum at the higher energies. In case {\bf D}, a jet
pointing towards the observer
without a speed restriction and with a faint disk and corona is
investigated. A high particle acceleration efficiency has also been considered for this particular
scenario, which shows that the jet could be a very strong emitter of X-rays and gamma-rays under
suitable conditions, appearing almost as an ultra luminous X-ray source (ULXs) if located at large distances. We
recall that it is supposed that all these systems are in a low-hard-like state, when compact jets are
thought to be present with rather stable characteristics (e.g. Gallo et~al. \cite{Gallo03}).

\section{The model: formation and properties of the jet}\label{jet}

We describe in this section the simple scenario we adopt concerning the formation, collimation and other
properties of the jet but the radiative ones, which are given in Sect.~\ref{rad}. The fixed parameters of
the model, their representative symbols, description and values are provided in Table~\ref{tab}, being 
discussed
in the text. For the stellar mass loss rate ($\dot{m}_{\rm w}$), the jet size, and the orbital
parameters, we have used typical MQ values. The parameter values that vary for the four specific scenarios
{\bf A}, {\bf B}, {\bf C} and {\bf D} are given below the fixed ones in Table~\ref{tab}.

\begin{table*}[]
  \caption[]{Model parameters}
  \label{tab}
  \begin{center}
  \begin{tabular}{lllll}
  \hline\noalign{\smallskip}
 Parameter: description [units] &  values \\
  \hline\noalign{\smallskip}
$e$: eccentricity & 0.3 \\
$a$: orbital semi-major axis [$R_{\odot}$] & 45 \\
$\dot{m}_{\rm w}$: stellar mass loss rate [$M_{\odot}$~yr$^{-1}$] & 10$^{-6}$ \\
$kT_{\rm disk}$: disk inner part temperature [keV] & 0.1 \\
$p_{\rm cor}$: corona photon index & 1.6 \\
$R_{\rm disk}$: disk inner radius [$R_{\rm Sch}$] & 50 \\
$r_{\rm l}$: launching radius [$R_{\rm Sch}$] & 4 \\
$z_0$: jet initial point in the compact object RF [$R_{\rm Sch}$] & 50 \\
$\chi$: jet semi-opening angle tangent & 0.1 \\
$\varrho$: equipartition parameter & 0.1 \\
$\zeta$: max. ratio hot to cold lepton number & 0.001 \\
$q_{\rm acc}$: fraction of the Edington accretion rate & 0.05 \\
$p$: electron power-law index & 2.2 \\
\hline\noalign{\smallskip}
\hline\noalign{\smallskip}
 & {\bf A} & {\bf B} & {\bf C} & {\bf D} \\
\hline\noalign{\smallskip}
$M_{\rm x}$: compact object mass [$M_{\odot}$] & 3 & 3 &  3 & 15 \\
$R_{\star}$: stellar radius [$R_{\odot}$] & 15 & 1 & 10 & 10 \\
$M_{\star}$: stellar mass [$M_{\odot}$] & 30 & 1 & 20 & 20 \\
$L_{\star}$: stellar bolometric luminosity [erg~s$^{-1}$] & 10$^{39}$ & 10$^{33}$ & 10$^{38}$ & 10$^{38}$ \\
$T_{\star}$: stellar surface temperature [K] & $4\times10^4$ & $6\times10^3$ & $3\times10^4$ & $3\times10^4$ \\
$\kappa$: jet-accretion rate parameter & 0.05 & 0.05 & 0.02 & 0.01 \\
$\xi$: shock energy dissipation efficiency & 0.25 & 0.25 & 0.05 & 0.5 \\
$\theta$: jet viewing angle [$^{\circ}$] & 45 & 45 & 45 & 1 \\
$\eta$: acceleration efficiency & 0.1 & 0.1 & 0.0001 & 0.1 \\
$\alpha_{\rm disk}$: disk radiative efficiency & 0.025 & 0.025 & 0.25 & 0.025 \\
$\alpha_{\rm cor}$: corona radiative efficiency & 0.005 & 0.005 & 0.05 & 0.005 \\
  \noalign{\smallskip}\hline
  \end{tabular}
  \end{center}
\end{table*}

\subsection{Jet formation}\label{jetf}

\subsubsection{Stellar matter accretion}\label{vel}

The accretion rate adopted in this work has been estimated assuming that the system is accreting at 5\% of the
Eddington value, thought to be typical for accreting XRBs in the low-hard state (Esin et~al. \cite{Esin97}). For the
accretion luminosity to accretion rate ratio, we have adopted 0.05$c^2$, which for a compact object of 3~$M_{\odot}$
yields an accretion rate of $6\times10^{-9}$~M$_{\odot}$~yr$^{-1}$. Assuming a spherical Bondi-Hoyle accretion model 
(Bondi \cite{Bondi52}, see also Reig et~al. \cite{Reig03}), in the case of high-mass MQs, with typical stellar mass
loss rates of about $10^{-6}$~M$_{\odot}$~yr$^{-1}$ and high wind velocities of few times 10$^8$~cm~s$^{-1}$ inferred
from spectroscopic observations, the mentioned moderate accretion rate is hard to achieve. It likely implies that
some  anisotropy in the wind properties occurs in the direction towards the compact object. We have assumed a low wind
velocity in the direction of the compact object that can be estimated from the stellar mass loss and accretion rates
quoted above (see also Table \ref{tab}), obtaining values of $\la 10^8$~cm~s$^{-1}$. Moreover, the adoption of an
accretion model allows us to provide a rough estimate of the effects of eccentric orbital motion on emission
variability as presented in Sect.~\ref{seds}, normalizing the accretion rate to the quoted value at phase 0. We adopt
the convention here that phase 0 is the periastron passage and when the compact object is in opposition to the
observer. We note that periastron passage does not correspond to the accretion peak, due to effects concerning the
composition of velocities for both the compact object and the wind. In the case of low-mass MQs, orbital variability is
not explored here because these systems present circular orbits.

\subsubsection{Ejection velocity of the jet}\label{vel}

The jet velocity is estimated taking into account the amount of available kinetic luminosity that
can be extracted from the accretion at the launching radius $r_{\rm l}$
(see Table~\ref{tab} and Sect.~\ref{disc}). $r_{\rm l}$ is not the distance 
at which the jet is formed, but a characteristic radius where ejected matter gets extra kinetic energy 
from the accretion reservoir. The knowledge of such a quantity could help us to understand where ejection 
originates, although ejection is likely to be a spatially extended phenomenon.
It is necessary to establish the matter
rates of both the advected and the ejected matter components that, in relation to the total accretion
rate, follow the formula:
\begin{equation}
\dot{m}_{\rm acc}=2\dot{m}_{\rm jet}+\dot{m}_{\rm adv}.
\label{eq:accmat}
\end{equation}
The factor 2 is due to the existence of a jet and its counterjet. We
introduce the parameter $\kappa$ through the relationship: $\dot{m}_{\rm jet}=\kappa \dot{m}_{\rm acc}$. 
To assign a certain amount
of extra kinetic luminosity to the ejected matter, we have taken into account the energy dissipated
in the disk and the corona in the form of radiation, as well as the energy borne by the advected matter
after transferring part of its kinetic energy to the jet. This 
remaining
advected kinetic luminosity has been assumed to be associated with the Keplerian velocity at the 
launching 
radius, which is: 
\begin{equation}
L_{\rm k\; adv}(r_{\rm l})\sim\frac{R_{\rm Sch}}{4r_{\rm l}}\dot{m}_{\rm adv}c^2.
\label{eq:adven}
\end{equation}
The previous considerations give a first order estimate of the injection jet velocity. The
equation we obtain is:
\begin{equation}
L_{\rm acc}(r_{\rm l})=L_{\rm k \;jet}+L_{\rm k\; adv}+L_{\rm disk}+L_{\rm cor}.
\label{eq:jetvel}
\end{equation}
$L_{\rm k\; jet}$ is the jet kinetic luminosity at the launching radius. This 
accounts for the kinetic luminosity required to carry the jet matter outside the potential well 
(e.g. it could be in the form of magnetic luminosity in those regions, before the jet is completely 
formed at the assumed distance of 50~$R_{\rm Sch}$) 
plus the kinetic luminosity of the jet after ejection:
\begin{equation}
L_{\rm k\; jet}(r_{\rm l})=\frac{GM_{\star}2\dot{m}_{\rm jet}}{r_{\rm l}}
+(\Gamma_{\rm jet}-1)(2\dot{m}_{\rm jet})c^2.
\label{eq:pot}
\end{equation}
$(\Gamma_{\rm jet}-1)(2\dot{m}_{\rm jet})c^2$, the jet matter kinetic luminosity, accounts also for 
the magnetic field and relativistic particle luminosity since, once formed, the jet is assumed to be cold matter 
dominated.
$L_{\rm disk}$ is the disk radiated luminosity, taken to be a few \% of 
$L_{\rm acc}$, and $L_{\rm cor}$ is the corona radiated luminosity, 
about 1\% or less of $L_{\rm acc}$ (the efficiencies that we have adopted are similar to 
those found in the literature; McClintock \& Remillard \cite{McClintock04}): 
$L_{\rm disk}=\alpha_{\rm disk} L_{\rm acc}$ and $L_{\rm cor}=\alpha_{\rm cor} L_{\rm acc}$.
The final expression for the $\Gamma_{\rm jet}$ is:
\begin{equation}
\Gamma_{\rm jet}(r_{\rm l})=1+\frac{1}{2\dot{m}_{\rm jet}}\left(\frac{R_{\rm Sch}}{4r_{\rm l}}\dot{m}_{\rm adv}-
\frac{L_{\rm disk}+L_{\rm cor}}{c^2}\right).
\label{eq:gjet}
\end{equation}

We note that the energy dissipated in the shocks formed in  the jet is extracted from the corresponding jet
kinetic luminosity after ejection, and the velocity at infinity has an associated kinetic 
luminosity such that $L_{\rm k\; \infty}<L_{\rm k\; esc}$.  Therefore, the ejection jet Lorentz 
factor will be larger than its observable value, $\Gamma_{\infty}$, although both will be related by 
the amount of energy dissipated in the jet (see below). This approach is classical, 
just to provide a zeroth order estimate.

In Sect.~\ref{seds} the computed SEDs for four different cases are presented. Except in the case of {\bf D},
$\Gamma_{\rm jet}$ is fixed to 1.5. This will allow us to compare, through Eq.~(\ref{eq:gjet}), the amount of
matter carried by the jet  between the case when disk/corona emission is weak and the case when it is strong. This
will imply that, for the same ejection velocity, the jet in the former case can be heavier than in the
latter one (see the corresponding $\kappa$ values in Table~\ref{tab}).

\subsubsection{Magnetic field}\label{mag}

Although we do not consider any particular theory of jet ejection, it is supposed that the
mechanism is a magneto-hydrodynamic one. If the jet is ejected by converting magnetic energy to matter
kinetic energy, it seems likely that $B$ close to the compact object must be beyond
equipartition with jet matter. We assume that $B$ goes down as it transfers 
energy to the jet matter, accelerating it. Finally, $B$ becomes dynamically dominated by jet matter (in the jet 
reference frame (RF)). 
It is assumed that, when the jet is already formed, the magnetic field is entangled with
matter and approximately turbulent. Both matter and $B$ follow adiabatic evolution when moving 
along the jet, with the energy density $\propto 1/z^2$ for a conical jet, where $z$ is the distance
to the compact object.

The transition from a magnetic field dominated jet to a matter
dominated jet should not be discontinuous; $B$ must reach values below equipartition in relation to matter, 
but not by too much, since their energy densities
should evolve in a similar way once the former is attached to the latter. 
A more extended discussion of this issue concerning
extragalactic jets can be found in Sikora et~al. (\cite{Sikora05}). In our work,  the magnetic
field in the jet reference frame at different distances from the compact object has been
calculated as follows:
\begin{equation}
B(z)=\sqrt{\varrho \;8\pi e_{\rm p}},
\label{eq:mag}
\end{equation}
where, for a cold proton dominated jet, the jet matter energy density is:
\begin{equation}
e_{\rm p}(z)=\frac{\dot{m}_{\rm jet}}{\pi R_{\rm jet}^2 V_{\rm jet}m_{\rm p}}<E_{\rm p\; k}>=
\frac{\dot{m}_{\rm jet}}{2\pi z^2}V_{\rm jet}.
\label{eq:mag}
\end{equation}
We introduce $R_{\rm jet}=\chi z$, and $<E_{\rm p\; k}>$, which is the mean cold proton kinetic
energy, taken to be the classical kinetic energy of protons with velocity equal to the
expansion velocity ($V_{\rm exp}=\chi V_{\rm jet}$). This would correspond approximately 
to the sound speed of the plasma in the jet RF.

\subsection{Jet properties}\label{content}

\subsubsection{Confinement}\label{conf}

\begin{figure}
\resizebox{\hsize}{!}{\includegraphics{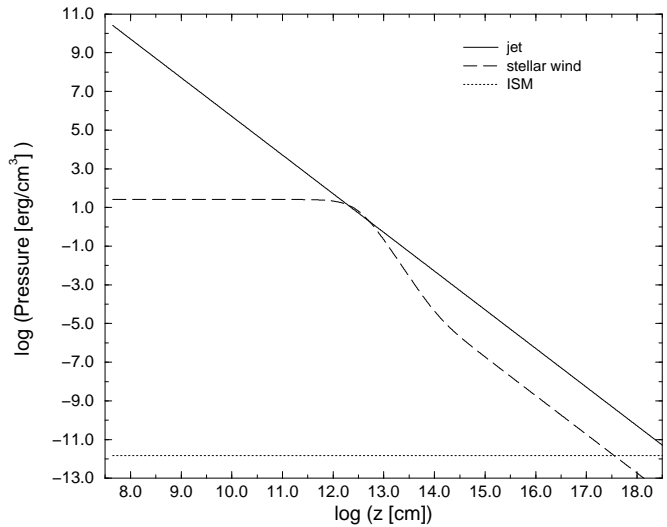}}
\caption{Pressure values along the jet for the jet matter (solid line), stellar wind (dashed line) 
and the ISM (dotted line). To compute this, $\kappa=0.05$ and a ISM with 
a density of 10~cm$^{-3}$ and a temperature of 10$^3$~K have been adopted.}
\label{press}
\end{figure}

Compact jets in microquasars appear to be in general mildly relativistic (for LS~5039, see Paredes et~al.
\cite{Paredes02};  for LS~I~+61~303, see Massi et~al. \cite{Massi04}; in general, see Gallo et~al.
\cite{Gallo03}). This implies that these jets cannot be collimated by relativistic motion, and an
external or internal factor must collimate them.  External medium collimation operates when its pressure is
similar to or larger than the jet lateral pressure, both taken as a non relativistic ideal gas (the jet
expansion velocity considered here, as well as the environment gas, is not relativistic). Environment pressure
can be exerted by expelled stellar matter (e.g. stellar wind) and the interstellar medium (ISM). In
Fig.~\ref{press}, we show the jet pressure along the jet, the stellar wind pressure and the ISM pressure. The
jet specific parameter values relevant here are those corresponding to the case {\bf A}. For the ISM, we have
adopted a density of 10~cm$^{-3}$ and a temperature of 10$^3$~K, assuming that it is heated by the massive star.
For this particular case, the wind velocity has been fixed to 10$^8$~cm~s$^{-1}$. As is seen in
Fig.~\ref{press}, the pressure of the wind is not significantly above the jet pressure anywhere along the jet. 
There is a clear change of behavior of the wind pressure on the jet when it turns from kinetic to thermal pressure,
at distances similar to the binary system size. We note that ISM effects could be non-negligible at very large
scales. We have not studied the interaction of the jet with the ISM, although a treatment of this kind can be
found elsewhere, e.g., in Heinz (\cite{Heinz02}). Further sources of external influence could  be disk radiation
and/or disk winds. However, the former does not appear to be effective at middle and large jet scales (see, e.g.
Fukue et~al. \cite{Fukue01}), and the latter, if effective beyond disk scale distances, would need to be as
compact as the jet itself, and the problem of collimation is transferred to this compact disk wind. 

Internal collimation could be achieved by a special magnetic field configuration. Nevertheless, since the
magnetic field  energy density is considered to be lower than the matter energy density (see Sect.~\ref{jetf}),
this mechanism does not seem to be relevant here. In our context, the jet expands freely at roughly its sound
speed, and this type of jet can only be collimated if the plasma sound speed is low, implying a cold matter
dominated jet. For instance, the jet of SS~433 is particularly heavy among MQ compact jets and seems to be cold
matter dominated, since there is significant thermal X-ray radiation coming from the jet (see, i.e., Fender
et~al. \cite{Fender03b}).  Jet opening angles are typically of a few degrees (for LS~5039, see Paredes et~al.
\cite{Paredes02};  for SS~433, see Marshall et~al. \cite{Marshall02}). This means that the parameter $\chi$ must
be similar to or less than 0.1 (see Table~\ref{tab}). For mildly relativistic jets, if
their cold components are in thermal equilibrium, the temperature associated with $V_{\rm exp}$ will be similar
to that of the inner regions of the  accretion disk or corona (several 100~keV). 

In order to obtain some upper limits for the amount of accelerated particles, we have computed both the cold and the
hot proton pressure along the jet. For hot particles, we have estimated their pressure as corresponding to a
relativistic ideal gas of protons following a power-law distribution. It is found that the pressure of the hot
component is similar to that of the cold component for a ratio of hot particle to cold particle number of about
1/1000. We have adopted this value for $\zeta$, the parameter that gives the maximum possible ratio of hot to cold
lepton number (we assume the existence of one cold lepton per cold proton). We note that the energy density of relativistic
leptons is below the magnetic one all along the jet. This fact could imply a better confinement for such
particles, considering that the gyro-radius becomes smaller as the magnetic field increases.

\subsubsection{Relativistic particles}\label{relpart}

Internal shocks due to different velocities of the jet plasma  (Rees \cite{Rees78}, Spada et~al. \cite{Spada01},
Yuan et~al. \cite{Yuan05})  can dissipate bulk kinetic energy, converting it into random kinetic energy of
accelerated particles. The injected particle spectrum is kept all along the jet up to a certain maximum energy,
which varies in accordance with the balance of energy gains and losses. 
In addition, the assumption that  the jet is cold matter
dominated stringently constrains $\zeta$ to the quoted value. 
Effects of cooling on the power-law
spectrum  cannot be considered at this stage due to the uncertainties concerning the injection/acceleration
processes, and its shape  index is therefore fixed. Accelerated leptons emitting synchrotron radiation  can
produce jet emission similar to that observed at radio wavelengths in LS~5039 (Paredes et~al.
\cite{Paredes00}),  the microquasar detected by HESS. 

Under the
conditions assumed here, the first order Fermi mechanism cannot accelerate thermal electrons (Bell \cite{Bell78})
efficiently. Therefore, a still unspecified mechanism that can accelerate thermal electrons must operate up to
the minimum  Fermi acceleration energy, $\gamma_{\rm min}$. To determine $\gamma_{\rm min}$, we have assumed
that the accelerated leptons have a mean free path within the shock similar to the mean free path of the
particles that form the shock, i.e. the jet cold protons. The injection mechanism remains unspecified (e.g.
magnetic reconnection that heats electrons, some type of proton-electron temperature coupling, injection of
relativistic pairs, etc), and the injection rate will depend on acceleration efficiency constraints. 
The Fermi mechanism provides the energy of radiating leptons, and not this unknown
injection mechanism. Thus, the radiative properties do not rely on a mysterious process of particle injection
but on the well-known first order Fermi theory  of particle acceleration (a similar approach to hadron
acceleration is performed in Mastichiadis \& Kirk \cite{Mastichiadis95}).  

The relativistic lepton energy distribution in the RF of the jet is assumed to follow a
power-law:
\begin{equation}
n_{\rm e}(\gamma,z)=\frac{N_{\rm inj}}{\pi (\chi z)^2 V_{\rm jet}}\gamma^{-p},
\label{eq:eldist}
\end{equation}
where $N_{\rm inj}$ is the injection normalization parameter for the energy distribution of relativistic 
leptons,
which is taken to vary as a function of the acceleration mechanism efficiency and jet conditions (see 
below). The injected lepton rate $\dot{Q}_{\rm inj}$ (i.e. the total number of leptons crossing a jet 
section at $z$ per second) is associated with $N_{\rm inj}$, and can 
be obtained from the latter integrating $N_{\rm inj}\gamma^{-p}$ from $\gamma_{\rm min}$ to 
$\gamma_{\rm max}$.
$p$ is taken to be 2.2, a reasonable value that can be derived from optically thin 
radio spectra observed in some MQs.
Since the matter density decays as $1/z^2$ in a conical jet, we observe the z-dependence in Eq.~(\ref{eq:eldist}). 
The acceleration process is assumed to keep 
the same energy distribution for the relativistic leptons along the jet,
although the maximum energy of the accelerated particles depends on energy loss conditions and the size of the 
accelerator, taken here to be the jet width. 
We will neglect at this stage the effects of escaping particles on the particle energy distribution
(for a detailed treatment on this see 
Atoyan \& Aharonian \cite{Atoyan99}). The power-law is simply cut at a certain $\gamma_{\rm max}$,
which is computed as explained below. Nevertheless, since the accelerator/jet size limits the 
acceleration efficiency, escape losses are taken into account as cooling terms. 
We note that a low energy particle population could be present in the
jet just below the acceleration injection energy, that can have Lorentz factors $\la$100.
These particles cannot radiate significantly, due to their long radiative timescales and because of adiabatic
losses, at least while they are confined within the jet. 

\subsubsection{Particle acceleration and jet deceleration}\label{partacc}

We have adopted Fermi first order acceleration theory to calculate the maximum energy of accelerated
particles in the jet (see, e.g., Biermann \& Sttritmatter \cite{Biermann87}, Protheroe
\cite{Protheroe99}). We set the acceleration rate equal to the rate at which particles
lose their energy. Expressed in terms of the Lorentz factor: 
\begin{equation}
\dot{\gamma}_{\rm gain}=\dot{\gamma}_{\rm loss}.
\label{eq:accel}
\end{equation}
The energy gain rate can be calculated from:
\begin{equation}
\dot{\gamma}_{\rm gain}=\frac{\eta q_{\rm e}Bc}{m_{\rm e}c^2},
\label{eq:accener}
\end{equation}
where $\eta$ can have different values depending on the shock
conditions\footnote{It depends on the angle between the magnetic field lines 
and the perpendicular direction to the shock surface, and on the shock speed in the plasma frame, as well as the 
diffusion coefficient in the shock region. $\eta$ is 
typically between $\sim 10^{-4}-10^{-1}$ (Protheroe \cite{Protheroe99})}. 
For simplicity and since the specific shock conditions 
are not known, $\eta$ is treated here as a free parameter in the range 10$^{-4}$--10$^{-1}$.
$\eta$ is likely a function of $z$, 
although at this stage is set to be constant. 
We recall that $B$ goes down by $1/z$.
The energy loss rate can be estimated adding the 
contribution of the different types of energy losses. On the one hand, there are adiabatic losses:
\begin{equation}
\dot{\gamma}_{\rm adiab}=\frac{2V_{\rm exp}\gamma}{3R_{\rm jet}},
\label{eq:lossadiab}
\end{equation}
since relativistic leptons are exerting work against the jet confining walls, whose expansion is led by cold
protons. On the other hand, there are radiative losses due to synchrotron, IC and relativistic
Bremsstrahlung processes. Hence, we obtain:
\begin{equation}
\dot{\gamma}_{\rm loss}=\dot{\gamma}_{\rm rad}+\dot{\gamma}_{\rm adiab}.
\label{eq:lossener}
\end{equation}
The expressions for the particle energy loss rates for each radiative mechanism can be found, e.g.,  in
Blumenthal \& Gould (\cite{Blumenthal70}). From these expressions together with Eq.~(\ref{eq:accel}), one can
finally obtain $\gamma_{\rm max}$ and its evolution with $z$. 

There is a limit for the total amount of energy available for shock acceleration, which is likely related to the
shock efficiency to dissipate energy via heating jet leptons. The upper limit for shock energy dissipation
efficiency is that 100\% of the kinetic energy is dissipated. It seems unlikely that a so efficient process
takes place but, still, an important amount of the energy is released in the form of radiation, as it appears to
happen for extragalactic jets and some galactic jets.  We have adopted the criterion that, at most, the shock
could dissipate up to some fraction of the flow average kinetic energy accelerating particles in the whole jet,  or 
$\xi=0.05, 0.25, 0.5$ (see Table.\ref{tab}). 
It is not stated that a 5, 25 or 50\% 
of the whole jet kinetic energy goes to the relativistic particles. The
shock could dissipate enough energy to accelerate the maximum number of relativistic leptons in the jet (given
by $\zeta$), still being below the efficiency limit.  This implies that for most of the jet the energy
dissipation rate can be below this constraint. If a large fraction of the jet kinetic energy  goes to heat
particles, the shock dynamics are affected. We do not include such effects in our calculations, treating the
acceleration process as the test particle case. We do not include the effects of the shocks in the local
conditions either, but one can consider them as averaged ones (for more precise calculations, properties of the 
shocks like velocity or compression ratio would be required, which is beyond present knowledge and
it is absorbed by other parameters like $\eta$ and $\varrho$). $\xi$ might be interpreted as referring to a
feedback effect on shock dynamics produced by accelerated particles.

The next constraint has been imposed:
\begin{equation}
\frac{dL_{\rm dis}(z)}{dz}\ge\frac{dL_{\rm loss}(z)}{dz},
\label{eq:dis}
\end{equation}
where $dL_{\rm loss}(z)/dz$ is the energy lost by length unit at different $z$ through 
radiative and adiabatic losses and has the form:
\begin{equation}
\frac{dL_{\rm loss}(z)}{dz}=\int^{\gamma_{\rm max}(z)}_{\gamma_{\rm min}}\pi R_{\rm jet}^2
n_{\rm e}(\gamma,z)
\dot{\gamma}_{\rm loss}d\gamma,
\label{eq:emiss}
\end{equation}
provided that all the quantities are already in the compact object RF. $dL_{\rm dis}(z)/dz$ 
is the maximum dissipated luminosity per length unit, which is taken to evolve with $z$ like 
$\dot{\gamma}_{\rm gain}$:
\begin{equation}
\frac{dL_{\rm dis}(z)}{dz}\sim \frac{C_{\rm dis}}{z}.
\label{eq:dis}
\end{equation}
The normalization constant $C_{\rm dis}$ can be obtained by integrating the total amount of shock 
energy available per length unit along the jet, from $z_0$ to $z_{\rm max}$, and equating it to a
suitable fraction ($\xi$) of the jet kinetic luminosity:
\begin{equation}
C_{\rm dis}=\frac{\xi L_{\rm k~esc}}{\ln(z_{\rm max}/z_0)}.
\label{eq:ctdis}
\end{equation}
$dL_{\rm dis}(z)/dz$ has a weak $z_{\rm max}$-dependence. This quantity has
been taken to be about 0.1~pc, where environmental effects on the jet properties could become significant, 
changing the characteristics of the jet, as it has been argued in Sect.~\ref{content}. 

\begin{figure}
\resizebox{\hsize}{!}{\includegraphics{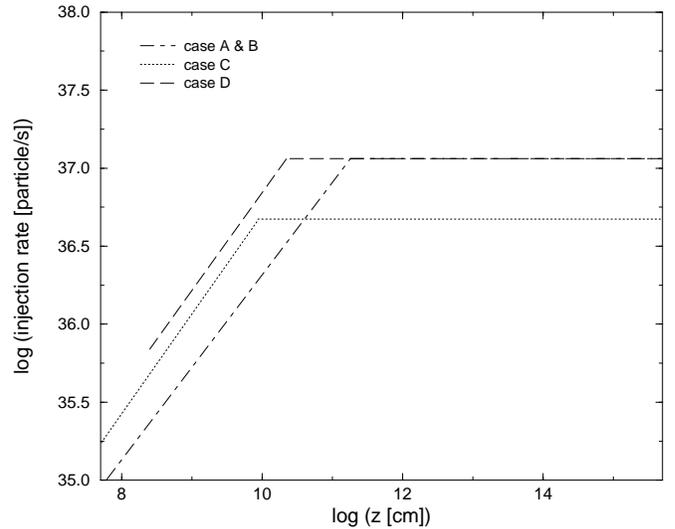}}
\caption{The evolution of $\dot{Q}_{\rm inj}$ along the jet for {\bf A}, {\bf B}, 
{\bf C} and {\bf D}. 
The free parameter values adopted for these cases are shown in Table~\ref{tab}.}
\label{ninj}
\end{figure}

Equating Eq.~(\ref{eq:emiss}) and Eq.~(\ref{eq:dis}) (through Eq.~\ref{eq:eldist}) gives $N_{\rm inj}$ in the compact 
object RF, although
this value cannot be higher than that derived from $\zeta$ either. 
In  Fig.~\ref{ninj}, we show
the evolution of $\dot{Q}_{\rm inj}$ for the four MQ cases studied here. Losses are
strong enough well within the binary system to hold the total number of relativistic leptons injected per second 
(i.e. crossing a jet section at $z$)
below the maximum allowed rate. As these particles are carried by the jet to larger $z$, losses are lower
and the injected particle rate reaches its maximum value. 
For the cases considered here, synchrotron  losses dominate over SSC, EC and
relativistic Bremsstrahlung losses, although star IC losses can dominate at certain jet regions and/or
for slightly smaller values of $B$. Therefore, the Klein-Nishina (KN) effects are to be taken into account. 

\begin{figure}
\resizebox{\hsize}{!}{\includegraphics{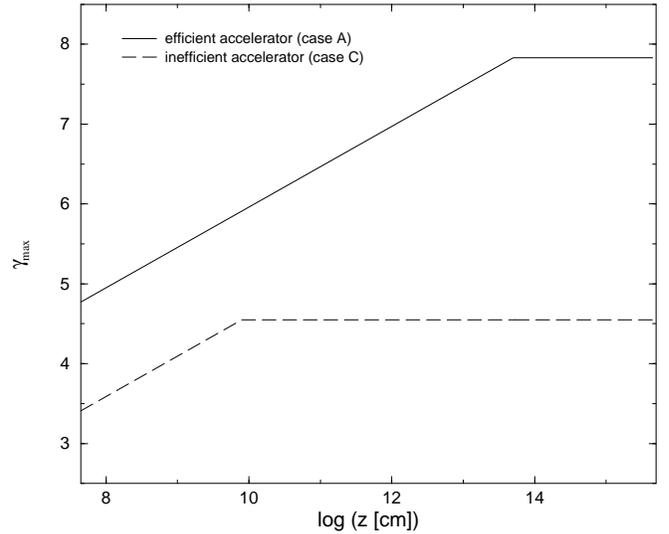}}
\caption{The same as in Fig.~\ref{ninj} but for $\gamma_{\rm max}$ in the cases {\bf A} and {\bf C}.}
\label{gammax}
\end{figure}

\begin{figure}
\resizebox{\hsize}{!}{\includegraphics{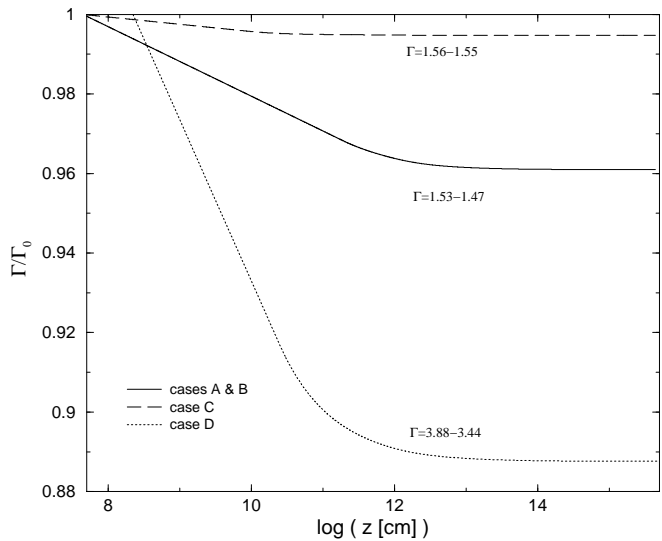}}
\caption{The same as in Fig.~\ref{ninj}, but for $\Gamma_{\rm jet}$ normalized to its initial value
in the four cases explored here.}
\label{gamjet}
\end{figure}

In Fig.~\ref{gammax}, we see the evolution of the maximum energy of the relativistic particles with $z$.  
Radiative losses limit this maximum energy at spatial scales similar to the size of the binary system. 
At middle and large jet scales, the
acceleration process is limited by the typical size of the jet and by the jet magnetic field. Both 
$z$-dependences get canceled, and the consequence is a constant value for $\gamma_{\rm max}$. We show the
evolution  of $\gamma_{\rm max}$ for two particularly interesting cases (Fig.~\ref{gammax}): {\bf A} and {\bf
C}. Since the amount of energy transferred to relativistic particles can be significant, we
calculate the effects of this process on the jet Lorentz factor. Since different velocities 
in plasma motion generate 
shocks and the acceleration process, the jet Lorentz factor must be understood as an average
value. The results
for cases {\bf A}, {\bf B}, {\bf C} and {\bf D} are shown in Fig.~\ref{gamjet}. As is 
expected, the lower the acceleration efficiency (i.e. the transfer of energy from the bulk motion to the
radiating particles), the lower the decrease in the bulk motion Lorentz factor, which goes down strongly at the
base of the jet and is already stabilized at binary system scales. The radiative efficiency of the jets for our
particular parameter choice (for $\varrho$,  $\xi$ and $\eta$) is of about 1--10\%, not far from
estimates obtained by different  approaches (see, e.g., Fender \cite{Fender01}, Yuan et~al. \cite{Yuan05}). 
With the Lorentz factors given in Fig.~\ref{gamjet}, the initial jet kinetic luminosities are $8.7\times10^{36}$~erg~s$^{-1}$, 
$8.7\times10^{36}$~erg~s$^{-1}$, $3.5\times10^{36}$~erg~s$^{-1}$ and $4.9\times10^{37}$~erg~s$^{-1}$
for {\bf A}, {\bf B}, {\bf C} and {\bf D} respectively.

\section{The model: radiative processes in the jet}\label{rad}

\subsection{Radiation mechanisms}

\begin{figure}
\resizebox{\hsize}{!}{\includegraphics{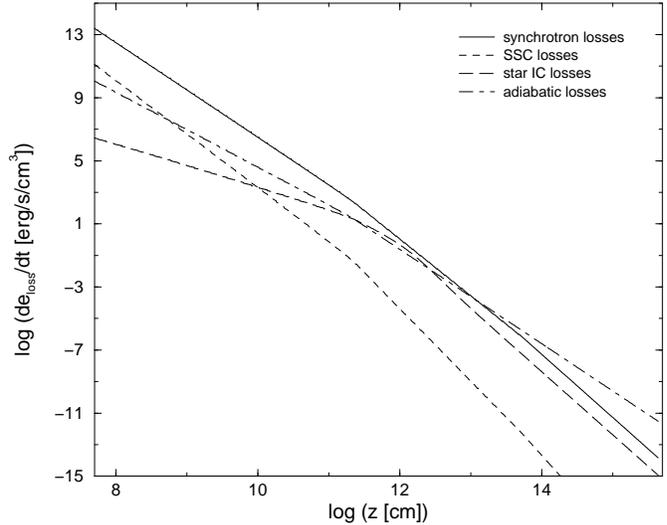}}
\caption{The same as in Fig.~\ref{ninj} but for the energy per volume and time units lost by leptons, 
$de_{\rm loss}/dt$, now only for {\bf A}.}
\label{lossrad}
\end{figure}

We have accounted for synchrotron, relativistic Bremsstrahlung and IC emission (in both Thomson and KN regimes).
Other leptonic radiative processes such as free-free transitions, thermal Bremsstrahlung, etc, have been
neglected since jet emission is dominated by the non-thermal processes, although for a completely cold and
powerful jet these processes would have observable effects.We show in Fig.~\ref{lossrad} an example of the
energy lost by leptons per volume and time units in the jet at different distances from the compact object. As
is seen there, the dominant type of loss is synchrotron radiation, although the IC losses can be significant.
Well outside the binary system, the dominant losses are the adiabatic ones. 
For certain ranges of the parameter values, stellar IC ({\bf A}-like but with low magnetic field) and corona IC
({\bf C}) could dominate over synchrotron radiation.  Relativistic Bremsstrahlung is negligible in
general. Since the increase of the amount of relativistic 
particles stops when their number reaches a maximum value (recall $\zeta$ and Fig.~\ref{ninj}), 
the decrease in the total emissivity becomes more serious 
at distances of about 10$^{11}$~cm from the compact object.

Taking into account the physical conditions along the jet,  we have computed the SEDs corresponding to the different 
mechanisms mentioned above. To calculate the spectrum of the radiation coming from the jet,  we have used the energy
distribution function of the relativistic leptons shown in Sect.~\ref{content}, as well as the corresponding
cross-sections of each process.  For synchrotron emission, the magnetic field at different $z$ is given in
Sect.~\ref{jetf}, and we adopt the width of the jet as the length that determines whether synchrotron emission is
either optically thin or optically thick.  For external Bremsstrahlung, in the case of interaction with the stellar
wind ions (the wind is considered as a completely  ionized plasma), we take our calculations as an upper limit and
assume that the target ion density is the wind density (i.e. that all the available wind particles diffuse within the
jet); for internal Bremsstrahlung, the target ion density is the proton density of the jet, directly derived from
$\dot{m}_{\rm jet}$ since the jet is cold matter dominated.  For EC emission, the target photon densities from  the
star, the disk and the corona are those described in Sect.~\ref{extph}, and in the case of internal Compton emission
(basically, SSC), the target photon density has been calculated previously (i.e. synchrotron).  All the $z$-dependences
have been  taken into account, dividing the jet in slices, each with homogeneous properties, and the overall jet
emission has been integrated over all the slices. For further details concerning synchrotron, relativistic
Bremsstrahlung and IC processes, as well as the electron energy loss expressions, we refer to the work by Blumenthal \&
Gould (\cite{Blumenthal70}). The Doppler boosting effects in the observed spectra have been implemented as usual (e.g.
Dermer \& Schlickeiser \cite{Dermer02}).

\subsection{External photon fields} \label{extph}

The external seed photon sources considered in the model are the star, the disk and the corona.  The star
and the disk have been considered to be gray bodies, normalized to their total luminosities. The star
photon distribution peaks at optical-UV energies, and the disk one peaks at 100~eV.  This disk photon
energy appears to be typical in the  low-hard state when the optically thick disk is truncated far from the
compact object and the disk matter does not reach  temperatures as high as during more intense accretion
states, when the inner disk radius shrinks.
For the 
corona emission, we have assumed that it follows a power-law plus an exponential cut-off, peaking
around 100~keV. Since IC interactions are studied first in the jet RF, we calculate the total energy
densities and the spectral energy densities in this RF. Since expressions for the star, disk and
corona photon energy density are originally in the compact object RF, they are transformed 
using a relationship found elsewhere (e.g. Dermer \& Schlickeiser \cite{Dermer02}):
\begin{equation}
U_{\epsilon_0,\Omega}(z)=\frac{U_{\epsilon_0',\Omega'}'(z')}{\Gamma_{\rm jet}^3(1+\beta\mu)^3},
\label{eq:star} 
\end{equation} 
where $\Omega$ represents the photon direction,
$\mu$ is $\cos\vartheta$, $\vartheta$ is the angle between the photon direction and the jet axis, and 
$\epsilon_0$ is the seed photon energy for IC interaction. 
The quantities with primes are in the compact object/observer RF, and in the jet RF otherwise.
For the star photons we assume that they reach the jet, at a particular $z'$, with the same direction, which depends on the orbital phase and $z'$. In the case of disk photons, 
their direction is taken to be coming from {\it behind} the jet.
To treat mono-directional seed photon fields, we have adopted the approach used by Dermer
et~al. (\cite{Dermer92})\footnote{For the mildly relativistic jets treated here, the disk IC radiation 
coming from the counterjet is enhanced by the angular dependence of the IC interaction. 
However, for the adopted parameter values, this component is at most similar to the {\it normal} jet 
component.}. Since the interaction angle between jet electrons at different
$z'$ and stellar photons affects the star IC emission and varies with the orbital phase, 
we recall the adopted criterion that the systems treated here are
observed in such a way that at phase 0 or periastron passage, the compact object is in opposition 
to the observer. To apply the model to particular objects, the specific relationship between 
observer line of sight and compact object phase is needed. For corona photons,
the jet base in this model is assumed to be located approximately in the external parts of this region. 
This source
of photons is closer to the jet than the disk, and the interaction is significant 
only at the inner region of the 
jet due to the $1/z^2$-decrease in density of corona photons. 
Thus, we have adopted the assumption that the corona field is roughly isotropic. 

\subsection{Pair creation and annihilation within the jet}\label{pairs}

\begin{figure}
\resizebox{\hsize}{!}{\includegraphics{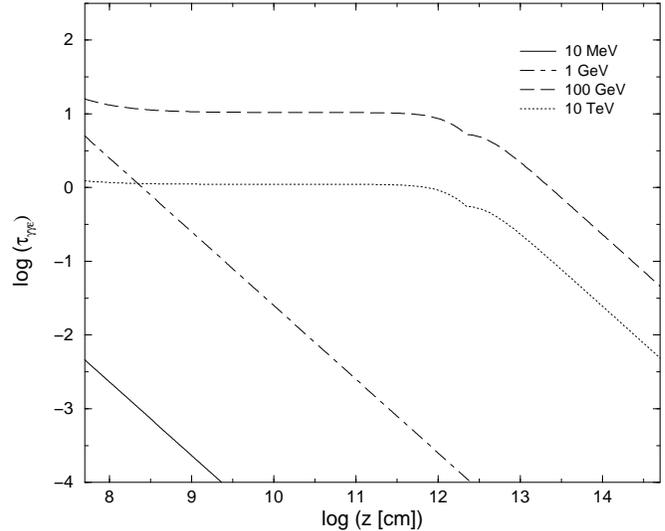}}
\caption{Evolution of $\tau_{\gamma\gamma\epsilon'}$ at different $z'$. We have adopted 
the parameter values of the case {\bf A} (see Table~\ref{tab}).}
\label{abs}
\end{figure}

We have investigated the effects of pair creation and annihilation phenomena on the jet emission by calculating
the gamma-ray opacities ($\tau_{\gamma\gamma\epsilon'}$) for jet photons of different $\epsilon'$ and produced
at different $z'$. To calculate opacities as well as the number of created and annihilated pairs, we have used
the pair creation and annihilation rates given in Eqs.~3.7 and 4.6 of Coppi \& Blandford (\cite{Coppi90}). The
evolution of the opacity is shown in Fig.~\ref{abs}. It is seen that opacities are higher in the vicinity of the
compact object at energies between 1~GeV--10~TeV. Gamma-ray absorption in the stellar UV photon field  is
significant for photon energies of $\sim100$~GeV (for a deeper treatment of gamma-ray opacities due to the
stellar photon field, see Romero et al. \cite{Romero05}, B\"ottcher \& Dermer \cite{Bottcher05} and Dubus
\cite{Dubus05}). For certain parameter values, the opacity could be significant as well at 10~MeV and a few GeV
within the corona and disk fields (see also Romero et al. \cite{Romero02}).

The gamma-ray opacity by pair creation inside the jet is very high at $z'\sim z_0'$ because the target
photon density is large and very sensitive to the jet width. If the jet width were larger, the internal 
opacity would be weaker, without affecting very much the overall production spectrum. Due to these 
uncertainties, the pair creation due to jet internal fields should be studied in a more detailed model in 
future work. 

Concerning annihilation rates inside this cold matter dominated jet, for any
reasonable set of parameter values the luminosity that could be emitted in form of an annihilation line is
too low to be distinguished from the continuum emission. Other models, like the one of Punsly et~al.
(\cite{Punsly00}), where a pure pair plasma is assumed, could produce detectable annihilation lines. 

Observable predictions from considering pair creation phenomena in our model are presented and discussed briefly
in Sect.~\ref{seds} and \ref{disc}, although we remark that the creation of pairs inside the jet could lead to
the appearance of bumps due to the accumulation of particles at the energies of pair creation. To introduce such
an effect properly requires a better knowledge of the particle injection function, which is beyond the scope of
this work.  Therefore, the high-energy gamma-ray band of the computed SEDs probably gives good
enough flux estimates, although slopes could be slightly different as a result of all these subtle effects. 

For those pairs that are created within the binary system, but outside the jet, the situation is different
from that of pairs created inside. Starting with a determinate number of relativistic particles in the jet,
plus the given jet conditions, one can consistently derive the SED of the produced radiation in the compact
object RF. Thus, the spectrum is known, and it allows us to know precisely the number of absorbed photons
and created pairs within the star, the disk and corona photon fields (for previous treatments of this, see
Romero et~al. \cite{Romero02}). From the previous result, it is possible to roughly estimate the radiation
that is generated by those pairs through IC interaction with external source photons. Although it is a
rough estimate, it is found to be in agreement with more detailed models of these processes (Khangulyan \&
Aharonian \cite{Khangulyan05}). 

\section{Application of the model} \label{seds}

The different radiation components produced in the jet and the predicted SEDs have been computed for the
four specific scenarios considered here. The effects of pair creation phenomena due to the external photon
fields interacting with the produced gamma-ray photons are taken into account, and the secondary radiation
produced by the created pairs is estimated.  The calculations are performed at the periastron passage, when
the compact object is in opposition to the observer and the interaction angle between star photons and jet
leptons implies more luminosity for the star IC component (see Dermer et~al. \cite{Dermer92}), 
showing the importance of such an effect. However, such an angle depends on the
electron energy, which should be taken into  account in more detailed models of the IC interaction 
(e.g., Khangulyan \& Aharonian \cite{Khangulyan05}).

\begin{figure}
\resizebox{\hsize}{!}{\includegraphics{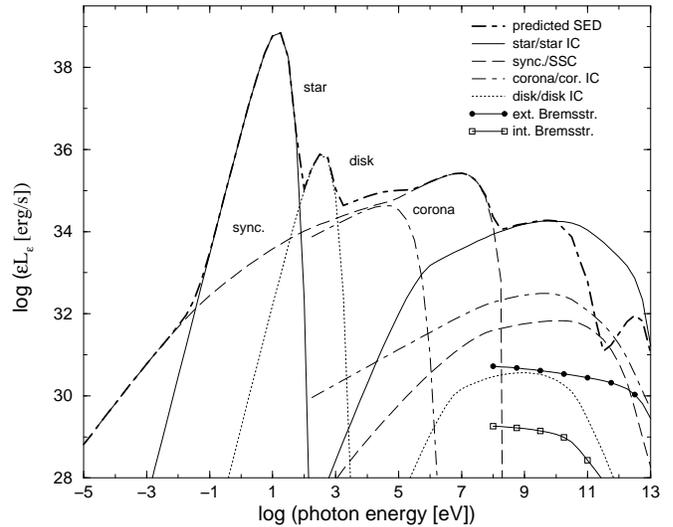}}
\caption{Case {\bf A} computed SED for the entire spectrum as it would be observed. Attenuation  of the jet photons due to
absorption in the external photon fields is taken into account, as well as the IC emission of the first generation of
pairs created within them. Isotropic luminosity is  assumed. The different IC, relativistic Bremsstrahlung, synchrotron
and other seed photon fields are shown. For the several components, the production SED is shown.  The corona photon field is also taken into account, but its effects on pair
creation  and subsequent emission are overcome by the synchrotron emission.}
\label{vhe}
\end{figure}

\begin{figure}
\resizebox{\hsize}{!}{\includegraphics{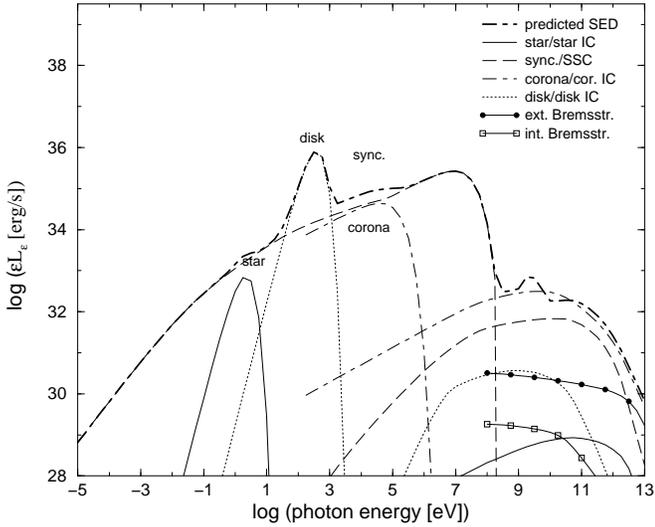}}
\caption{The same as in Fig.~\ref{vhe} but for the case {\bf B}.
The small bumps
present from beyond 100~MeV come from  the IC radiation emitted by those  leptons generated by pair creation in the
disk and the  stellar photon field. These pair components are not made explicit
in the  plot for clarity.}
\label{low}
\end{figure}

The broad-band SEDs for cases {\bf A} and {\bf B} are presented in Figs.~\ref{vhe} and \ref{low} respectively.  The
strong effects on the computed SED due to the presence of a massive star can be appreciated.  The star IC component is
very significant, partially because of the specific interaction angle between seed photons and leptons at phase 0, and
also because the interaction with stellar photons is more significant at large $z$, when the
number of relativistic particles is higher (see Fig.~\ref{ninj}), 
than for other photon fields. For {\it A}, gamma-gamma opacity is very high at VHE.
We recall that the disk and corona emission have been assumed to radiate just a few per cent of the  accretion power.
As accretion does not dissipate a significant fraction of the available energy via either disk or corona radiation, the
jet can carry more energy and matter for the same ejection velocity (and the assumptions put
forward in Sect.~\ref{jetf} are valid). The acceleration efficiency has been assumed to be high.

\begin{figure}
\resizebox{\hsize}{!}{\includegraphics{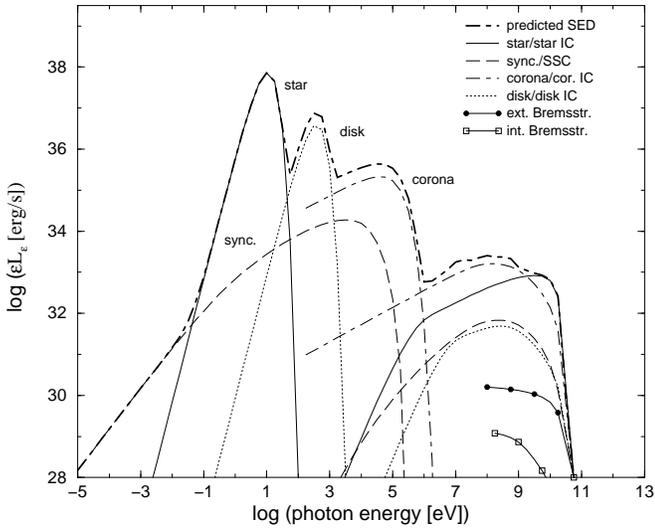}}
\caption{The same as in Fig.~\ref{vhe} but for the case {\bf C} with a jet 
with low acceleration efficiency. For this particular situation it is possible to see the small
bump at 10~MeV produced through IC scattering by the pairs created in the corona photon field.}
\label{xrb}
\end{figure}

\begin{figure}
\resizebox{\hsize}{!}{\includegraphics{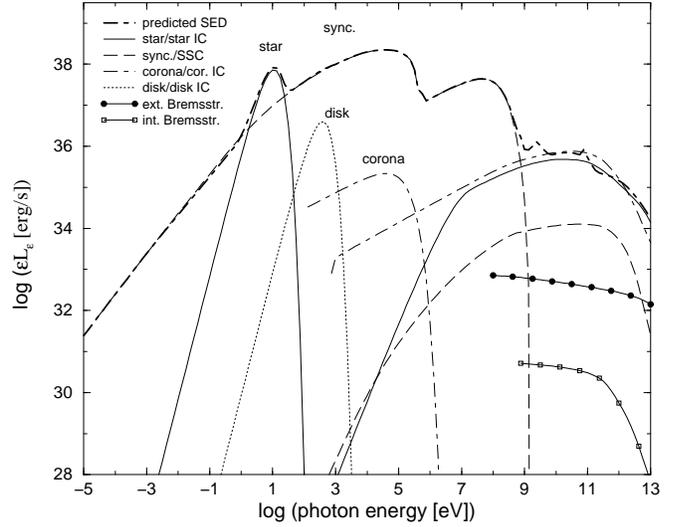}}
\caption{The same as in Fig.~\ref{vhe} but for the case {\bf D}.}
\label{ulxs}
\end{figure}

In Fig.~\ref{xrb}, the broad-band SED of case {\bf C} is shown. We have increased the disk and the corona
emission, fixing the jet velocity. This implies a lighter jet than in the two previous cases. Also, we
have modified the acceleration efficiency of the jet for this particular case to a smaller value than those
previously used. The system harbors a massive star, although neither as massive nor as bright as in {\bf
A}.  In the case of {\bf D}, whose SED is plotted in Fig.~\ref{ulxs}, the matter content of the jet has
been reduced to a smaller value, fixing again disk and corona emission and particle acceleration efficiency
as in {\bf A} and {\bf B}.  This implies that the jet motion will be more relativistic than in previous
cases. This is unusual in what has been said in this work. Our purpose in studying this particular
scenario is to show that an ultra luminous X-ray source can be reproduced through strong beaming and small
viewing angles. Since the goal in this specific scenario was to obtain very high X-ray fluxes, the compact
object has been considered a 15~M$_{\odot}$ black hole, allowing for higher accretion rates since the
Eddington limit is higher for more massive accreting objects.  We have also
increased the shock maximum 
energy dissipation efficiency up to $\xi=0.5$.

The SEDs obtained at high energies resemble roughly those obtained applying the model
presented in Bosch-Ramon et~al. (\cite{Bosch05a}). This is because the IC emission dominates at high energies and, at
least in some of the cases, the dominant IC component is the same as in that model. However, the
physical motivation of the present work goes much further than before, with predictions concerning radio, variability
and jet physical properties that could not be provided in the previous model.

\subsection{Spectral properties}

At gamma-rays, a different component is the main one in each explored scenario. For {\bf A}, the star
Comptonized photons are dominant, reaching $10^{34}$~erg~s$^{-1}$ at 100~MeV, with a photon index of about 2,
and $5\times10^{32}$~erg~s$^{-1}$ at 100~GeV. At VHE, due to the gamma-gamma absorption, the model predicts a
soft photon index that hardens at higher energies. There could be additional spectral features related to the
secondaries created in the corona, the disk and the star photon field. These features would
appear as bumps at  energies of a few GeV for disk pairs and a few tens of GeV for star photons 
(negligigle in {\it A}).  Even for
negligible disk and corona sources, a decrease in the predicted flux beyond 50~GeV is unavoidable if it is a
relatively close and massive binary system. In the case of {\bf B}, the dominant components are the corona IC
and the SSC ones, with luminosities of about $10^{33}$~erg~s$^{-1}$ at 100~MeV with a photon index of 2 (beyond
the synchrotron component, see Fig.~\ref{low}), and few $10^{32}$~erg~s$^{-1}$ at 100~GeV. In this case,
however, the low mass star photon field does not significantly affect the VHE spectrum, and only a small bump
due to the secondaries created within the disk  photon field is visible in the SED. Therefore, the photon index
beyond 10~GeV gets softer because of the KN effect, although it is a smoother steepening than in {\bf A}.
Concerning {\bf C}, the luminosity dominated by the corona IC emission at 100~MeV is about
$10^{33}$~erg~s$^{-1}$ and the photon index is similar to the one predicted in the previous cases. Otherwise,
only disk and corona gamma-gamma absorption effects are significant, visible in the plots at about 1~GeV and
10~MeV respectively (see Fig.~\ref{xrb}). The star IC component turns to dominates beyond a few GeV. For {\bf D},
the Doppler boosting affects synchrotron radiation, which reaches 100~MeV with luminosities higher than
$10^{37}$~erg~s$^{-1}$. However, the SSC component is not dominant at all because it is very sensitive to the jet
density, and the jet now is
the least dense among the four studied cases. Therefore, beyond 1~GeV the source is
dominated by the corona and star IC components, with a photon index softer than 2 and some absorption produced
by the star photon field.  Small peaks of the IC emission of the secondaries are also present produced within
the star and the disk photon fields.

In X-rays, the emission is synchrotron dominated for most cases (except for {\bf C}). This is because the
magnetic field is below but not far from equipartition. The matter energy density at the jet base is so high
that the magnetic field reaches values around 10$^5$~G, allowing synchrotron radiation to dominate up to soft
gamma-ray energies. Similar results have been obtained by Markoff et~al. (\cite{Markoff01},
\cite{Markoff03}) for the case of shock acceleration limited mainly by synchrotron losses. In our model particle
acceleration is limited by shock energy dissipation  efficiency, jet size, adiabatic and all the radiative
losses (including KN regime for IC losses). Because of the evolution of $\gamma_{\rm max}$ that rises when
$z$ is larger (for reasonable parameter values), the synchrotron spectrum changes smoothly at energies around
100~keV. Nevertheless, the disk and corona are not negligible in general, and in the case of {\bf C}, disk and corona
overcome the jet radiation up to 1~MeV. {\bf D} as an X-ray source is extremely bright.
This result shows that a light and fast jet observed from very small viewing angles might turn out to be an
ULX. Because of the low probability of being observed, it is more likely to detect them in other galaxies
(for previous works on MQs as ULXs, see, e.g., Georganopoulos et~al. \cite{Georganopoulos02}, K\"ording et~al. \cite{Kording02}).

Below stellar emission energies, synchrotron radiation dominates again. If stellar emission were reprocessed by
absorption and shifted to lower energies, it is likely that the far infrared would still be dominated by the enshrouded
stellar component. At radio frequencies, there is significant radio emission, with isotropic luminosities of about
10$^{29}$~erg~s$^{-1}$ at 5~GHz (few mJy at 3~kpc). Our model predicts core-dominated emission, strongly
self-absorbed due to the high efficiency of the inner jet to radiate through synchrotron process. The SED in the radio
band corresponds to a spectral index equal to zero (or $\epsilon L_{\epsilon}\propto \epsilon^{+1}$), as it would be
expected from a conical jet. {\bf C} presents the weakest radio emitting jet, weaker than in {\bf A} and {\bf B}, and
{\bf D} is the strongest radio source due to the Doppler boosting. Further aspects on radio emission in our scenario
are commented in next section.

\subsection{Variability}

\begin{figure}
\resizebox{\hsize}{!}{\includegraphics{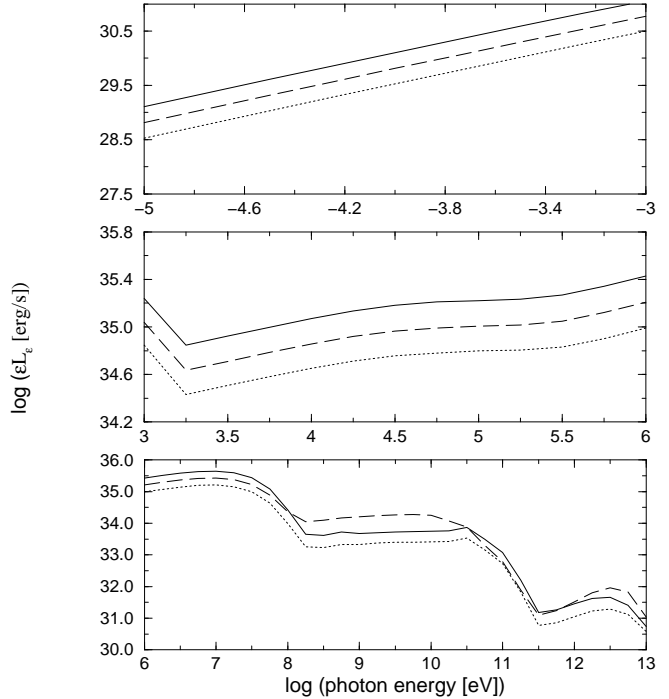}}
\caption{The case {\bf A}, three predicted SEDs at radio, X-ray and gamma-ray energy bands, corresponding to 
three different orbital phases: 0 (periastron passage, dashed line), 
0.3 (highest accretion rate, solid line), and 0.7 (smallest accretion 
rate, dotted line).}
\label{var1}
\end{figure}

Variability through changes in the stellar mass-loss density profile is introduced in the model in a consistent
way when referring to high-mass microquasars. In Fig.~\ref{var1}, the SEDs of {\bf A} at phases 0.3 (accretion
maximum), 0 (periastron passage) and 0.7 (accretion minimum) are shown. For clarity, we have split
the overall spectrum in three bands: radio, X-rays and gamma-rays. Emission varies due to orbital eccentricity
for a spherical slow wind accreting system with constant wind velocity. Even for the simple model considered
here, it is possible to see from the figures the complex evolution of radiation. At radio and X-ray frequencies,
the different fluxes associated with different phases are correlated with the accretion rate. Since the 
velocity of the compact object relative to its surrounding medium changes along the orbit, the
periastron is not directly associated with the highest flux. However, at gamma-rays, the particular angle of
interaction between star photons and jet electrons implies that the stellar IC is dominant, this phase being
associated with a peak in gamma-ray emission. More complex variations in the accretion, affecting jet
ejection and radiation, could be introduced through the function reproducing the stellar mass-loss density
profile. Other timescales of variability, linked to disk phenomena, could be introduced through the
parameter that controls the amount of matter that goes to the jet, although it is beyond the scope of this work.
Also, the radio variability pattern depends on the scale of the radio emitting region, which is the inner jet for
the parameter values adopted here. 

\section{Discussion} \label{disc}

We conclude that persistent jets in MQs in the low-hard state, despite accreting at relatively low rate, and under
reasonable conditions for the jet matter, energy and magnetic field, can radiate with significant luminosities from
radio to gamma-rays. The model provides  predictions about the shape of the SEDs and points to MQs as VHE sources, as
has been recently confirmed  by Aharonian et~al. (\cite{Aharonian05}) in the case of LS~5039.  It also predicts
variability at different ranges of energy. The importance of the synchrotron cooling channel for relativistic particles
in the jet is high. This points to the fact that, for certain objects, the jet could overcome at all wavelengths any
other emitting region of the MQ, except the star itself for high mass systems. We note that for a relatively weak disk
and corona  ($10^{34}$~erg~s$^{-1}$), and even with $\varrho\sim$0.01 (one tenth of the one adopted here), the
spectrum would be jet dominated and well described by a power-law at X-ray wavelegnths (see also Paredes et~al.
\cite{Paredes05}).  An observational feature that could determine whether X-rays come from a jet is the dependence of
X-ray fluxes on the accretion rate, if the latter can be estimated. In the scenario presented here, X-rays vary with
$\dot{m}_{\rm acc}$, but it is also indirectly connected to $B$ and $\gamma_{\rm max}$, the latter also being sensitive
to the overall jet conditions and size, and both quantities depend on the accretion rate as well. Long exposure
multiwavelength observations will be required if a relationship between radiation components of different origin is to
be found with high accuracy, particularly considering the moderate X-ray fluxes of permanent jet sources like LS~5039,
which do not appear to follow the {\it typical} behavior of X-ray binaries in the low-hard state (Bosch-Ramon et~al.
\cite{Bosch05c}). An additional observational feature that would favor the jet as the origin of the X-rays would be the
detection of some amount of polarization in this radiation.

At gamma-rays, our jet model, more detailed and in accordance to the current knowledge of MQs than our previous
works, still shows that MQs could be behind some EGRET sources (likely those variable and located in the
galactic plane), and also predicts a different evolution of emission at different energies due to IC angular
dependence interaction, some strange features like radiation bumps and depressions in the spectra due to
gamma-gamma absorption with external photon fields, and a non trivial relationship between pair creation and
particle injection within the jet itself.  GLAST and AGILE, with a sensitivity at least several times better
than that of EGRET, should be able to detect microquasars even when they are low mass systems (case {\bf B})
and/or they have weaker jets (case {\bf C}). The detection of the microquasar LS~5039 by HESS  shows that the
efficiency of the particle acceleration processes should be high.  Due to the strong photon absorption beyond
$\sim 100$~GeV, it might be that the bulk of the TeV emission came from regions where stellar photon density is
not significant.

Although we are using accretion rates which are a small fraction of the Eddington luminosity, the wind velocity is
required to be low to power accretion and the jet itself. This points to the fact that O stars with spherical fast
winds would not be able in general to power some of the compact jets observed in the galaxy, implying that some special
wind density profile should be given, likely produced by the presence of the compact object plus other factors like
companion star rotation, etc  (Paredes et~al. \cite{Paredes05}). For the launching radius, we have taken 4~$R_{\rm
Sch}$ to reach mildly jet Lorentz factors. Although our approach to estimate the energy balance between the jet and
accretion is rough, this value for $r_{\rm l}$ is between the last stable orbit and the limits of the corona-like
region, consistent with the present state of knowledge on this issue.

\subsection{Radio emission}

Previous models for leptonic emission from a magnetized compact jet predicted core dominated radio
emission as observed in several galactic and extragalactic sources (for XRBs jet models, see, e.g.
Markoff et~al. \cite{Markoff01}; for extragalactic jet models, see, e.g. Ghisellini et~al.
\cite{Ghisellini85}). Our results are similar to those presented in these previous works, where radio
emission that comes from the inner jet regions is strongly self-absorbed, becoming optically thin further down the jet. On the other hand, it is difficult to correctly model jet radio emission in some high-mass XRBs,
like that presented by LS~5039 (Mart\'{\i} et~al. \cite{Marti98}, Rib\'o et~al. \cite{Ribo99}). A deeper
discussion on this subject is presented in Paredes et~al. (\cite{Paredes05}). We advance however that a
$z$-dependence for the parameter $\eta$, or also a less stringent restriction of $\zeta$, could lead to a
higher production of radio emission in optically thin regions of the jet. 

A correlation is seen between the luminosity in the radio band and that at X-rays (Corbel et~al.
\cite{Corbel03},  Gallo et~al. \cite{Gallo03}) that appears to be present in different sources with very
different accretion rates and compact object mass values (Falcke et~al. \cite{Falcke04}).  In the case of our
model, when the source is jet dominated at X-ray wavelengths (cases {\bf A}, {\bf B} and {\bf D}),  the
correlation is fulfilled. Otherwise, for corona-dominated sources (case {\bf C}) the correlation cannot be
reproduced, since we are not modeling the corona. Instead, we adopt a typical spectrum and a certain luminosity
in each case.

\section{Summary} \label{sum}
 
We have developed a detailed leptonic model for an MQ jet that can reproduce the emission observed from radio to
gamma-ray energies, and makes precise predictions for very high energies. The basic assumptions of the model are
a cold-matter dominated jet, with a magnetic field close to but below equipartition that is entangled with and
dynamically dominated by jet cold matter.  With the knowledge of the system parameters, given a simple stellar
mass-loss density profile, and varying the jet to advected matter ratio and the acceleration efficiency, a set
of broad-band SEDs has been computed. Also, the opacity due to photon-photon interactions was taken into
account to calculate the predicted SEDs. The absorption can significantly distort the production spectrum beyond
10~GeV mainly due to the effect of stellar photons in the case of massive companion stars. The opacity can be
important even at lower energies when disk and  corona radiation densities are high enough. The model shows that
pair creation inside the jet could affect jet radiation. This aspect will be investigated
accurately in future work. For systems where orbital eccentricity or other stellar mass-loss asymmetries are
present, consistent predictions of the variability emission pattern of the source can be obtained. 

New generation gamma-ray instruments, both satellite-borne or ground-based, like GLAST, AGILE, MAGIC or HESS
can be used to test and constraint the high-energy predictions and assumptions of the model.  

\begin{acknowledgements}
We thank an anonynous referee for useful comments and suggestions that significantly 
improved the manuscript.
We are grateful to Evgeny Derishev, Dmitri Khangulyan, Felix Aharonian and John Kirk for useful discussions on
particle acceleration and plasma physics. V.B-R. and J.M.P. acknowledge partial support by DGI of the
Ministerio de Educaci\'on y Ciencia  (Spain) under grant AYA-2004-07171-C02-01, as well as additional
support from the European Regional Development Fund (ERDF/FEDER). During this work, V.B-R has been
supported by the DGI of the Ministerio de (Spain) under the fellowship BES-2002-2699. G.E.R is
supported by the Argentine Agencies CONICET and ANPCyT (PICT 03-13291).  This
research has benefited from a cooperation agreement supported by Fundaci\'on Antorchas.  \end{acknowledgements}

{}

\end{document}